\documentclass{article}
\begin{document}

\begin{center}
\centerline{\large \bf Time reversal noninvariance}
\centerline{\large \bf in quantum mechanics and in nonlinear optics.}
\end{center}

\vspace{3 pt}
\centerline{\sl V.A.Kuz'menko\footnote{Electronic 
address: kuzmenko@triniti.ru}}

\vspace{5 pt}
\centerline{\small \it Troitsk Institute for Innovation and Fusion 
Research,}
\centerline{\small \it Troitsk, Moscow region, 142190, Russian 
Federation.}
\vspace{5 pt}
\begin{abstract}

The experimental proofs of strong time invariance violation in optics are 
discussed. Time noninvariance is the only real physical base for explanation 
the origin of the most phenomena in nonlinear optics. The experimental study 
of forward and reversed transitions in oriented in uniform electric field 
molecules is proposed.

\vspace{5 pt}
\end{abstract}

\vspace{12 pt}

\centerline{\bf Introduction}

The idea of the arrow of time is quite natural and recognized in biology, 
chemistry and other fields [1]. However, its recognition in the physics 
unexpectedly turns out to be rather problematical because of the dynamical 
equations of the basic laws of physics are time reversal invariant. In 
classical theories the arrow of time is introduced as a result of time 
asymmetric boundary conditions with time symmetric dynamical equations 
[2, 3]. But for the full recognition of this concept the introduction of 
time asymmetry into the dynamical equations is a more preferable way [1]. 
And the nature gives us the grounds for such idea. However, this process, 
probably, will be rather long and hard [4, 5]. 

From our point of view the mistake of physicists is based on the two myths. 
The first myth is the opinion about absence of any experimental evidences 
of time invariance violation in electromagnetic interactions [1, 3, 6]. 
Below we shall discuss the three known for present day direct and 
independent experimental proofs of strong time invariance violation in 
optics. The second myth is the supposition about the existence of the 
so-called quantum coherent states [7]. This concept has now extremely wide 
spreading in physics. However, it turned out, that the concept of coherent 
states does not have any real physical base [8]. 

\vspace{5 pt}

\centerline{\bf First myth }

The interaction of polarized laser radiation with the specific non-magnetic 
metallic planar chiral nanostructures was studied in [9]. The authors 
believe that these experimental results unambiguously show the evidence of 
broken time reversal symmetry in such unusual object.

In other experiments the splitting and mixing of photons were studied
 [10, 11]. On the first stage the narrowband ($\sim 0,01 nm $) radiation of 
nanosecond laser was transformed through down-conversion in the nonlinear 
crystal into two intense broadband beams (each spectral width 
$ \sim 100 nm $ ). On the second stage this two broadband beams were mixed 
in the sum frequency generator [10] or in the process of two photon 
excitation of rubidium atoms [11]. The reversed process in these cases 
corresponds to mixing of the so-called entangled photons and leads to 
regeneration of the initial narrowband radiation. In contrast, the mixing of 
non-entangled photons should give broadband radiation and is the example of 
again only the forward process. Both experiments show the same results: the 
efficiency of reversed process is much greater, than the efficiency of 
forward process.

In the next case the forward and reversed transitions in $ SF_6 $ molecules 
were studied [12]. In those experiments the authors deal with such specific 
object as the so-called wide component of line in a spectrum of polyatomic 
molecules [13]. The forward transition ($CO_2$- laser photon absorption 
process) in this case has extremely high spectral width ($\sim 150 GHz$) 
and relatively small cross-section ($\sim 10^{-19} cm^2 $). In contrast, 
the reversed into the initial state transition (stimulated emission) 
has very small spectral width ($\sim 450 kHz$). The difference in spectral 
widths of forward and reversed transitions exceeds five orders of magnitude. 
Accordingly, the cross-section of the reversed process turns out to be in 
several orders of magnitude greater, than the cross-section of forward 
process. 

So, for the present day we have quite sufficient quantity of direct 
experimental proofs of a strong time invariance violation in optics. We have 
also enormous quantity of indirect experimental evidences, which are 
connected with the second myth about the existence of coherent states.

\vspace{5 pt}
	
\centerline{\bf Second myth }

The concept of coherent states appeared and became all pervasive in quantum 
optics during last several decades. The coherent states usually are 
interpreted as specific states of atoms or molecules after its interaction 
with a coherent laser radiation. Most phenomena in nonlinear optics are 
usually explained as a result of interference of the coherent states [15, 16]. 

However, the close theoretical analysis of this concept in [8] has shown 
that the inability to measure the absolute phase of an electromagnetic field 
prohibits the existence of quantum coherent states [7, 17]. It deprives 
the concept of coherent states any real physical sense. The concept of 
coherent states is only a "convenient fiction" for physicists in the field 
of quantum optics [8].

\vspace{5 pt}

\centerline{\bf Physical origin }

We believe that the only real physical base of observed phenomena in 
nonlinear optics is the time reversal noninvariance in electromagnetic 
interactions or inequality of forward and reversed processes in the optics. 
So, most phenomena, which are explained now as the interference of the 
coherent states, are really a manifestation of the time reversal 
noninvariance. However, time noninvariance manifests itself usually only 
in indirect way and this is the main reason why the concept of coherent 
states is popular till now.

What is the usual origin of inequality of forward and reversed processes 
in optics? The discussed above experiments show extremely high efficiency of 
the reversed process. However, for experiments with splitting and mixing of 
photons the origin of this efficiency is not clear as a whole. It is 
connected with the subtle concept of entanglement and it is the problem for 
future studies. 

For optical transitions in atoms and molecules the high efficiency of the 
reversed process is unambiguously connected with its high cross-section. 
However, we do not need to have a doubt in equality of the Einstein's 
coefficients for forward and reversed transitions [18]. The equality of 
Einstein's coefficients means only the equality of integral cross-sections 
of the opposite processes and does not prohibit the inequality of 
differential cross-sections. Such possibility is well illustrated by the 
discussed above experiments with $ SF_6$  molecules [12]. In this case the 
high cross-section of reversed optical transition is connected with its 
extremely small spectral width in contrast to the forward transition. 

The wide component of line is rather specific object and it exists only in 
the large polyatomic molecules. In small molecules and in atoms the line 
wings of such kind are absent. What is the reason in these cases of high 
differential cross-section of the reversed optical transition? 

Here we should pay attention to the quantum mechanical averaging process 
of some parameters of vibrational and rotational motions, which manifests 
itself in a molecule absorption spectrum. The lines in absorption spectrum 
of small molecules are very narrow. Its frequency allows calculating the 
moment of inertia of molecule with precision $\sim 0,0001 \%$. From other side 
the temporal change of moment of inertia during the period of vibrational 
motion of the atoms in molecule usually exceeds the value of $ 1\%$ [19]. 
This inevitably means that there exists some quantum mechanical process of 
averaging of moment of inertia during the period of vibration of the atoms 
in molecule. It is worth to mention that for polyatomic molecules this 
process of averaging, obviously, can undergo some short reversible 
violations even for the forward transition, which lead to appearance in 
absorption spectrum of the so-called line clumps [20]. 

The time asymmetry may consist in the difference between the averaging 
processes for forward and reversed transitions. For the forward transition 
the absorption cross-section does not depend from the phase of vibration 
motion. In contrast, for the reversed transition such dependence may exist. 
If the molecule has the phase of vibration motion, which allows it to return 
exactly into the initial state, then the backward transition will be reversed 
and its differential cross-section will be much higher than the averaged 
cross-section of the forward transition. In other cases the differential 
cross-section of the backward transition will be relatively small and such 
transition should be called over again only as the forward one. Inequality 
of forward and reversed processes in a natural way supposes the existence 
of the memory of atoms and molecules about the initial state. In some 
sense such memory can correspond to the entropy of quantum system.

So, the discussed above supposition is a good physical base for explanation 
of the origin of experimental study of the dynamics of vibration motion of 
atoms in molecule [21].

\vspace{5 pt}

\centerline{\bf Oriented molecules}

The similar quantum mechanical process of averaging of cross-section, 
probably, takes place for the rotational motion also. The rather common 
opinion exists, that the cross-section of interaction between the molecules 
and radiation should depend from orientation of molecules with respect 
to laser beam [22]. Some experiments with the anisotropy of fluorescence or 
with the successive transitions confirm this supposition. However, as a 
whole this is rather knotty problem. Practically all experiments in this 
field are carried out only with the linearly polarized laser radiation. 
Some authors even talk about the absorption of a linearly polarized photon 
[23]. Such photon, of course, is absent in nature. The linearly polarized 
light is a complex object and consists of the equal quantity of photons with 
different spins. So, we believe, that for the first stage of experiments 
with oriented molecules the more simple circularly polarized light should be 
preferably used. 

The process of orientation averaging of absorption cross-section is well 
illustrated, for example, by the experimental results of work [24], 
where the absorption of laser radiation by the hydrogen cyanide trimer 
molecules in the static electric field was studied. The external static 
electric field interacts with the dipole moment of molecule and tries to 
orientate it in the space. The experiments show, that when the energy of 
this interaction is smaller, than the energy of rotation motion, the 
perturbation of rotation motion practically does not manifest itself in 
the absorption spectrum. In the opposite case the molecules turn into 
the so-called pendular states and the absorption spectrum dramatically 
changes. 

The inequality of forward and reversed transitions, again, may be the result 
of different character of quantum mechanical processes of orientational 
averaging of cross-section of molecule interaction with laser field. 
For simplified illustration the Fig.1 shows the assumed dependence 
of transition's cross-section from the angle between the molecule axis and 
the direction of laser beam for forward (1) and reversed (2) transitions. 
The integral cross-sections of both transitions are equal. If the molecule 
is oriented in the space so that the backward process returns it exactly 
into the initial state, then the differential cross-section will be very 
large and such transition is reversed. In other orientation of molecule the 
backward process has relatively small differential cross-section and it 
should be called again only as the forward transition. In this case the 
molecule remains the memory about its initial state. So, in such way the 
origin of the so-called rotational coherency may be explained [25]. 

The other indirect evidence of inequality of forward and reversed transitions 
is the numerous experiments with degenerated or nondegenerated four photon 
mixing in the so-called folded boxcars arrangement [21]. Here the three 
laser beams of different directions are crossed in common point in a liquid, 
gas or in a molecular beam. The appearance of the superfluorescence in the 
new direction is observed. The cross-section of transition for such 
superfluorescence should be extraordinarily high. The direction and 
temporal characteristics of appearance of superfluorescence correspond to 
such orientation and phase of vibrational motion, which allow molecules to 
return exactly into the initial states. 

New and detailed information about the orientational dependence of 
cross-sections of forward and reversed transitions may, probably, be 
obtained in the experiments on the existent apparatus, which uses the static 
electric fields and molecular beam with cryogenic bolometer [12, 24, 26]. 
The Fig.2 shows the simplified arrangement for suggested experiments. The 
radiation of infrared laser is split on the two beams (pump and probe). 
The interaction regions are placed between the flat electrodes, which have 
slits for laser radiation and can be independently rotated round the 
molecular beam. As the object for experimental study the stable linear 
molecules with large dipole moment and small rotational constant well suit. 
There are, for example, $ HCCCN $ or $ H(CC)_n X$ molecules, were n=1--3 
and X - is the halogen. The low rotational states of such molecules may be 
easily turned into the pendular states at a relatively weak electric field. 
For the reversed optical transition we can expect, that the experimental 
dependence of its cross-section from the angle between the pairs of electrodes 
will be rather similar to the dependence on Fig.5 in work [12]. As a whole, 
such experiments can give important information about the orientational 
dependence of cross-sections for forward and reversed optical transitions.

\vspace{5 pt}

\centerline{\bf Conclusion}

Time reversal noninvariance is the only real physical base for explanation 
the origin of the most phenomena in nonlinear optics. For present day we 
have quite sufficient quantity of the direct experimental proofs of the 
strong time invariance violation in optics. This is a good reason to 
introduce the time asymmetry into the dynamical equations of the basic 
laws of physics. The corresponding asymmetric equations for description 
the dynamics of optical transitions will substitute for the famous Bloch 
equations, which are widely and successfully used in optics now [15], 
but which does not have any clear physical sense. The experiments for the 
study of orientational inequality of forward and reversed transitions in 
molecules are proposed.

\end{document}